**Inhomogeneity-driven multiform Spontaneous Hall Effect in conventional and unconventional superconductors**

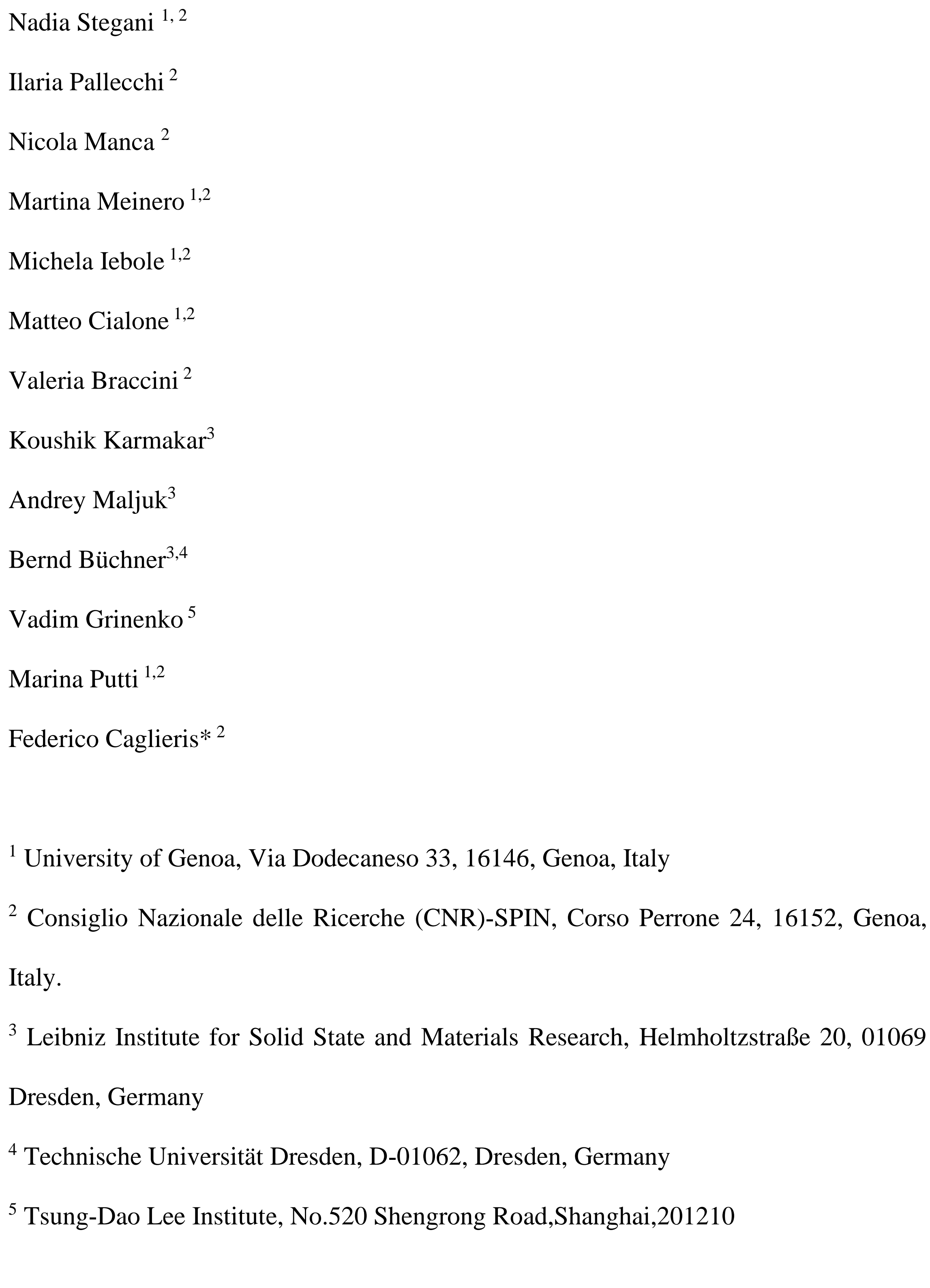

Nadia Stegani [1, 2]

Ilaria Pallecchi [2]

Nicola Manca [2]

Martina Meinero [1,2]

Michela Iebole [1,2]

Matteo Cialone [1,2]

Valeria Braccini [2]

Koushik Karmakar[3]

Andrey Maljuk[3]

Bernd Büchner[3,4]

Vadim Grinenko [5]

Marina Putti [1,2]

Federico Caglieris* [2]

[1] University of Genoa, Via Dodecaneso 33, 16146, Genoa, Italy

[2] Consiglio Nazionale delle Ricerche (CNR)-SPIN, Corso Perrone 24, 16152, Genoa, Italy.

[3] Leibniz Institute for Solid State and Materials Research, Helmholtzstraße 20, 01069 Dresden, Germany

[4] Technische Universität Dresden, D-01062, Dresden, Germany

[5] Tsung-Dao Lee Institute, No.520 Shengrong Road,Shanghai,201210

*federico.caglieris@spin.cnr.it

**Inhomogeneity-driven multiform Spontaneous Hall Effect in conventional and unconventional superconductors**

The spontaneous Hall effect (SHE), a finite voltage occurring transversal to the electrical current in zero-magnetic field, has been observed in both conventional and unconventional superconductors, appearing as a peak near the superconducting transition temperature. The origin of SHE is strongly debated, with proposed explanations ranging from intrinsic and extrinsic mechanisms such as spontaneous symmetry breaking and time-reversal symmetry breaking (BTRS), Abrikosov vortex motion, or extrinsic factors like material inhomogeneities, such as non-uniform critical temperature ($T_c$) distributions or structural asymmetries.

This work is an experimental study of the SHE in various superconducting materials. We focused on conventional, low-$T_c$, sharp transition Nb and unconventional, intermediate-Tc, smeared transition Fe(Se,Te). Our findings show distinct SHE peaks around the superconducting transition, with variations in height, sign and shape, indicating a possible common mechanism independent of the specific material.

We propose that spatial inhomogeneities in the critical temperature, caused by local chemical composition variations, disorder, or other forms of electronic spatial inhomogeneities could explain the appearing of the SHE. This hypothesis is supported by comprehensive finite elements simulations of randomly distributed Tc's by varying $T_c$-distribution, spatial scale of disorder and amplitude of the superconducting transition. The comparison between experimental results and simulations suggests a unified origin for the SHE in different superconductors, whereas different phenomenology can be explained in terms of amplitude of the transition temperature in respect to Tc-distribution.

## INTRODUCTION

One of the most intriguing areas of interest in conductive materials deals with their behaviour when subjected to a magnetic field. Within this domain, the Hall effect stands out as a well-known and significant phenomenon, describing the generation of a voltage transverse to the electrical current in applied perpendicular magnetic field, as result of the Lorentz force acting on charge carriers. The Hall effect plays a pivotal role in determining both the sign and density of charge carriers within a material, making it an indispensable tool for studying the properties of conducting materials.

If the electrical current is replaced by a thermal gradient as driving force, we enter the realm of the thermo-electric transport properties, where the counterpart of the Hall effect is known as the Nernst effect. In this scenario, we again observe the emergence of a voltage perpendicular to the direction of the heat gradient in a perpendicular applied magnetic field.

While the Hall and the Nernst effects typically vanish once the magnetic field is removed, it's noteworthy that a substantial number of studies deal with the occurrence of finite electric and thermoelectric transverse signals even in absence of external magnetic field. The most claimed examples are the so-called anomalous Hall (AHE) [1-5] and anomalous Nernst effects (ANE) [6-10], which often appear in ferromagnets and non-collinear anti-ferromagnets as a transverse voltage showing magnetic hysteresis, which remains finite when B=0. The AHE and the ANE can have an intrinsic origin, related to the Berry curvature associated to the charge carriers [11,12], or, alternatively, an extrinsic one determined by the presence of magnetic impurities, which causes skew scattering [13,14] or side-jump scattering of electrons [15].

The appearance of a finite transverse signal, even without a magnetic field, has been observed also in superconducting materials. Such phenomenon, renamed spontaneous Hall effect (SHE), appears in the form of a voltage peak across the superconducting transition, when the transverse resistivity is measured as a function of the temperature. In analogy, a spontaneous Nernst effect has been observed too.

The origin of the SHE in superconductors is strongly debated and different possible explanations have been proposed, ranging from intrinsic mechanisms related to spontaneous symmetry breakings, to the distribution of inhomogeneities inside the samples.

In particular, da Luz at al. invoked the Abrikosov vortices and anti-vortices motion to explain the spontaneous Hall effect in Pr-doped Bi2212 [16] and in polycrystalline YBCO [17]. More recently, a SHE has been suggested to be a signature of the breaking of time-reversal symmetry (BTRS) in a thin film of MgB2 [18], and of a spontaneous braking of inversion symmetry in Nb [19] or to be related to the topological structure of β-Bi2Pd [20]. In analogy, also the spontaneous Nernst effect has been put in relation with the BTRS in $Ba_{1-x}K_xFe_2As_2$ [21], FeSe [22] and $La_{2-x}Ba_xCuO_4$ [23].

Beside such intrinsic effects, an alternative explanation based on inhomogeneities has been also proposed since the 1990s [24,25]. Recently, Segal et al [26] justifies the existence of a spontaneous transverse voltage in Pb and YBCO by the presence of an asymmetric spatial inhomogeneity in the material, which causes a local transverse current component. The same idea has been previously suggested by Siebold et al. [27], who studied the current redistributions associated with the presence of Tc-inhomogeneities non-uniformly distributed, in superconductors. Finally, more recently, [28] transverse resistance measured on Nb and NbN thin films have been associated with the emergence of electronic spatial inhomogeneities.

The great number of studies regarding the SHE, encouraged us to begin an in-depth study of the occurrence of this phenomenon in a wide range of different superconducting materials. In Fig. 1 we present our data collection, showing the SHE peak (red dots) appearing around the superconducting transition of thin films of two conventional superconductors, Ta and Nb, as well as in a thin film of the unconventional iron-based superconductor $FeSe_{0.5}Te_{0.5}$ and in a bulk single crystal of the unconventional cuprate superconductor $La_{2-x}Sr_xCuO_4$ (LSCO). The superconducting transition (black dots) is superimposed to the peaks for each plot. The height of the peak varies between the samples, and the peak changes sign and shape.

Fig. 1 demonstrates that the SHE occurs in extremely different types of superconductors, belonging to different families, ranging from unconventional to BCS superconductors.

Given these relevant characteristics, it is plausible to consider that a common extrinsic explanation of the SHE, independent on the superconducting material taken into account, might exists.

In this work we propose a systematic and comparative study of the SHE. To this end, we selected thin films of two different superconductors: Nb, as a conventional, low-$T_c$ superconductor, and Fe(Se,Te) as an unconventional, intermediate-$T_c$ superconductor. We suggest a common explanation for the observed SHE based on the existence of a spatial inhomogeneity in the critical temperature within the sample, which can be either caused by a local variation of the chemical composition, strain, defects, dislocations or other form of disorder, or even electronic spatial inhomogeneities, like thermal/magnetic fluctuation, that may arise when adjacent or competing phases there exists [29, 30].

The comparison between experimental measurements and finite element simulations of the proposed scenario, corroborates our hypothesis, setting new boundaries to the interpretation of the emergence of the SHE.

## METHODS

***Thin film sample characteristics*:** The measurements presented in this work were performed on two thin films of Nb and Fe(Se,Te), with a nominal superconducting critical temperature Tc of 9K and 16K respectively. The films have a thickness of 80nm and 140nm respectively. Tc is extracted as the temperature value at which the normal state resistivity is reduced of 10%, while the superconducting transition width ΔTc is measured as the difference between the values at which the resistivity is reduced of 10% and 90%.

***Thin films growth***: The Fe(Se,Te) film has been prepared by Pulsed Laser Ablation technique on a Zr-doped CeO2 buffered YSZ substrate [31]. 20 nm thick Zr-doped CeO2 films were deposited via a Chemical Solution Deposition method known as Metal Organic Decomposition (MOD) on YSZ single crystals [32]. Fe(Se,Te) films were deposited on such template by pulsed laser deposition (PLD) in an ultra-high vacuum chamber (residual gas pressure during deposition of about $10^{-8}$ mbar during deposition) equipped with a Nd:YAG laser at 1024 nm from a polycrystalline target synthesized through a two-step method with a nominal composition $FeSe_{0.5}Te_{0.5}$ [33]. First, a 100 nm thick, non-superconducting seed layer was deposited at high temperature (about 400 °C) and at high laser repetition rate (10 Hz) [34]. Then, after cooling to 200 °C, a 100 nm film was grown at 3 Hz. The film was deposited having fixed the laser fluency at 2 J $cm^{-2}$ with a 2 $mm^2$ spot size and 5 cm distance between target and substrate.

The Nb sample has been prepared by R. Russo via vacuum arc technique in ultra-high vacuum conditions (UHVCA) [35,36] on a 2μm sapphire substrate. The deposition has been performed by applying an arc current of 120A with a $V_{bias}$=-40V. The total pressure during arc operation is in the $10^{-5}$ Pa range (10−7 mbar).

***Hall-bars micro-fabrication***: Superconducting thin films were patterned into an array of Hall-bars by standard UV lithography. SPR-220-4.5 photoresist was deposited by drop casting followed by spin-coating at 6000 RPM for 45 s. After a baking step at 120 °C for 135 s the device geometry was transferred by masked UV exposure, then the samples were put soaking in MR-D 526 until the developing was complete. Thin film etching was performed by Ar ion milling at 500 eV and 0.2 mA/cm² current density for about 10 mins. Finally, the samples were cleaned in ethanol bath and dried under nitrogen flow. The size of the Hall bars was 1mm x 0.5mm with an internal channel of 50 um in width.

***Electrical Measurement***: The electrical measurements have been performed in a Quantum Design PPMS equipped with a 9 T superconducting magnet. Electrical resistivity was measured in four-wire configuration on films micro-machined in Hall-bar geometry. The longitudinal and transverse signals, in zero magnetic field, are measured simultaneously. The SHE peaks have been extracted from the transverse and longitudinal resistivity curves, as explained in [37].

***Synthesis of the LSCO sample*****:** The LSCO single crystal used for the measurements presented in fig. 1d was grown by travelling solvent floating zone (TSFZ) method on the conventional image furnace (CSC, Japan) under $O_2$ pressure of 2 bars. The sample was annealed in $O_2$ flow at 450 C for 5 days. More details are given in Ref. [38].

## EXPERIMENTAL

The longitudinal resistivity and the SHE of Nb and Fe(Se,Te) thin films are reported in figure 2. The upper panels (a and c) show the longitudinal resistivity measured on two identical Hall bars, placed in different positions, for each sample, while the lower panels (b and d) present the SHE peaks. The superconducting parameters, critical temperature Tc and transition width ΔTc, of both samples have been measured, and are shown in fig. 2f.

The longitudinal resistance $\rho_{xx}$ of the Nb thin film indicates a superconducting transition around 9K, for both Hall bars. Hall bar 1 presents a very sharp transition, while in Hall bar 2 it is enlarged and presents a double transition feature at low temperature (see fig 2a). The corresponding SHE signals, presented in Fig 1b, show that the SHE peaks measured in the two Hall bars have different sign – positive for Hall bar 1 and negative for Hall bar 2 - and shape, and a "double peak" feature appears correspondingly to the double-transition behaviour detected in the longitudinal resistivity across Hall bar 2. However, the identical widths of the SHE peaks and of the superconducting transitions indicate that these two phenomena are intertwinned, despite the amplitude of SHE peaks is about ten times lower than the corresponding longitudinal signals.

Figure 2c presents the $\rho_{xx}$ curves of the two Hall bars measured on the Fe(Se,Te) thin film. The superconducting transition shows an onset at around 16K, but the transition width (ΔT) varies between the two Hall bars. For Hall Bar 1 ΔTc is of about 1K, while for Hall Bar 2 ΔTc is nearly double, as reported in fig. 2a. Similarly to the Nb case, the SHE peaks in Fe(Se,Te), reported in fig. 2d, are tightly related to their corresponding superconducting transition, in particular by matching its width, while the sign and the absolute value of the SHE peaks change from one Hall bar to the other. Finally, the amplitude of the SHE peaks in Fe(Se,Te) is 100 times smaller than the $\rho_{xx}$ amplitude.

Figures 3 shows the effect of an applied magnetic field up to 1T on Nb thin film and up to 9T on the Fe(Se,Te) thin film.

The magnetic field affects the superconducting transition of Nb (see fig. 3a) primarily by causing a rigid shift of Tc to lower temperatures. The upper critical field, $H_{c2}(T)$, extracted by the shift of Tc in field, is reported in fig. 3g. Moreover, a detailed analysis reveals a slightly widening of the width of the superconducting transition, which is especially noticeable at 1T. Interestingly the double transition feature in the Hall bar 2 disappears for magnetic fields higher than 0.25T. This indicates a different field dependence of the regions with different Tc detected by the double-peak feature. In particular, the low Tc region, responsible for the knee in the curve at 0T, seems to have steeper $H_{c2}(T)$.

The SHE peaks (fig. 3b) keep appearing in both Hall bar 1 and 2 for all the applied magnetic field intensities and their widths continue to reflect the width of the transition. Interestingly, the shape and the sign of the SHE peaks change, but their amplitude decreases with increasing the field. Crucially, the SHE peak is symmetric with respect to the applied magnetic field, which makes it to be clearly distinguished from an eventual standard Hall effect (see [37]).

The behaviour of the superconducting transition of Fe(Se,Te) in a magnetic field up to 9T for both Hall bars is presented in fig. 3c and fig. 3e respectively, and it mainly results in a broadening of the transition, while its onset just slightly shift to lower temperatures, thus indicating a high $H_{c2}(T)$ of this superconductor, as reported in figs. 3h and 3i. Again, the width of the SHE peak for all the magnetic field intensities (figs. 3d and 3f) follows the width of the corresponding transition, and the onset of the peak keeps fixed by increasing the field. In this case, a clear trend of progressive decrease in peak height with increasing applied magnetic field is easily observed.

**FINITE ELEMENTS SIMULATIONS**

The basic idea behind our work is that inhomogeneities within the sample are capable to produce a finite SHE. When we perform an electrical transport measurement in a non-magnetic material, we can assume that the current follows the injected direction only if the sample is perfectly homogeneous. In such a case, the current will not have any transverse component, and any measurement of transverse voltage will return a zero value in absence of applied magnetic field. However, if the sample is inhomogeneous, the current will seek the best percolative path, flowing through regions with lower resistance. As a result, the current direction will deviate from the straight path, creating local transverse current components and leading to the appearance of a transverse voltage. In superconducting materials, this effect is significantly amplified across the superconducting transition. In such transient phase, zero resistivity is not reached simultaneously in all the parts of the sample, and this causes a significant and abrupt change in the current density distribution. Therefore, across the transition, the transverse voltage will differ significantly from the longitudinal voltage, leading to the appearance of the SHE peak.

Our work involves reconstructing the phenomenon by finite elements simulations using COMSOL Multiphysics. Specifically, we considered a superconducting film of size 1.0 x 0.5 x 0.001 $mm^3$ (thickness << lateral size). The spatial distribution of disorder in the sample is simulated using a $T_c$ distribution centred around 20K across the sample plane and uniform along the out-of-plane direction. In correspondence of each $T_c$ distribution, also a resistivity distribution ensues. Specifically, for each local $T_c$ value, a local resistivity curve is assumed, with a transition width $\Delta T_c$. The mathematically modelled $T_c$ distribution is presented in [37]. In particular, we analyzed scales of disorder of 10, 15, 30, 70, 400 and 800 $\mu m$, ranging from values much smaller than the sample size to values comparable to the sample size. Fig. 4a shows, as example for the reader, the 2D maps of $T_c$ (upper panels) together with the corresponding maps of resistivity (lower panels) for disorder scales of 10, 30 and 400 μm.

Two values $\Delta T_c = 1$ K and $\Delta T_c$=0.1 K were considered. The former larger $\Delta T_c = 1$ K represents the transition broadening due to disorder on a much smaller scale, that replicates uniformly throughout the sample on the considered micrometric or larger scales, or alternatively it represents the transition broadening due to fluctuations induced by temperature or magnetic field. This situation corresponds to a superconducting material whose properties depend on its composition, such as doping or stoichiometry, as for Fe(Se,Te). The smaller $\Delta T_c$=0.1 K represents an ideally sharp transition with no disorder and no fluctuations. This is the case of a superconducting pure element, such as Nb. Fig 4b shows the effect of the local transition width $\Delta T_c$=1 K or $\Delta T_c$=0.1K on the ρ maps, for a disorder scale of 30μm, while the Tc maps is unchanged between the two cases. By narrowing $\Delta T_c$, the map of ρ changes significantly: this is because the narrower the transition, the more abruptly the resistivity varies when moving from one region of the sample to another. On the other hand, when ΔTc is larger, the variation of ρ is smoother due to an "averaging" effect between the regions. In other words, the narrower is ΔTc, the higher are the spatial frequencies of Tc variations that contribute to determine the local direction of the current.

Finally, three different configurations were simulated, by starting with different random seeds for each disorder scale, giving us crucial insights into how SHE depends not only on the spatial scale of disorder but also on its specific configuration. In fig 4c the three simulated configurations A, B and C, for the scale of disorder of 30 μm are shown as an example.

Figure 5 shows the longitudinal Vxx and transverse Vxy voltage drop as a function of temperature for all scales of disorder for the configuration of disorder B, fixing all the other parameters of the problem, both in the case of a local transition width of $\Delta T_c$=1 K (panels a and b), and in the case of $\Delta T_c$=0.1 K (panels c and d). Longitudinal and transverse voltages are detected simultaneously once a current of 1mA is applied. In case of $\Delta T_c$=0.1 K the curves related to the scales of disorder of 10 and 15μm are lacking since the simulation does not converge. Indeed, as the scale of disorder gets smaller and/or the $T_c$ variations get more abrupt, adjacent grid elements turn out to have larger mismatch in conductivity, which challenges the continuity conditions on current distribution and its derivatives, required as convergence constraint.

At a first sight, regardless of the scale of disorder and the transition width considered, an important characteristic of the SHE peak emerges from the simulations: the intensity of the SHE peak is at least one order of magnitude lower than the intensity of the corresponding Vxx, while the width of the peak matches the width of the corresponding superconducting transition. To support the last statement, we invite the reader to observe the insets of the figures 5a and 5c, where the feet of the longitudinal transitions are enlarged.

The transverse voltage in panels b and d exhibit the SHE peak at each scale of disorder with different sign and magnitude. A general trend can be identified independently of the transition width: the peak becomes narrower and smaller in magnitude with increasing spatial scale of disorder from 10 to 400 μm for the $\Delta T_c$=1 K case, and from 30 to 400 μm for the $\Delta T_c$=0.1 K case. Fig. 5f highlights this trend, showing the absolute value of the peak integral both for the simulation with $\Delta T_c$=1 K (black dots), and the simulation with $\Delta T_c$=0.1K (red dots). Eventually, this trend does not describe anymore the behavior of the peak when the spatial scale of disorder is comparable to the size of the sample (spatial scale of disorder 800 μm, as compared to sample planar size 1.0 x 0.5 $mm^2$). This trend, together with the 800μm exception, is well reproduced also for the configuration of disorder A and C, presented in fig. 6. However, this general trend is not universal, and it may be violated when the increasing (decreasing) spatial scale of disorder causes areas of high Tc or low Tc to merge (separate) on a size comparable to the sample width, thus closing (opening) transverse path of high conductivity at temperatures around $T_c$ (see the case of 10 μm spatial scale of disorder in Fig. 5b). Finally, it's important to highlight that the peak integral for the 30, 70 and 400μm disorder scales perfectly match between the $\Delta T_c$=0.1 K and the $\Delta T_c$=1 K case, indicating that as the width decreases, the height increases, maintaining a constant area. Indeed, a direct comparison between the simulations obtained with $\Delta T_c$=1 K and $\Delta T_c$=0.1 K (fig 5e) shows that a narrower transition width implies an increase in the height of the peaks.

Fig. 6 presents the longitudinal Vxx and transverse Vxy signals as a function of temperature for the other two configurations of disorder A and C, investigated in this work, both for the $\Delta T_c$=0.1K and $\Delta T_c$=1K case, while varying the disorder scale from 10 to 800 μm. For both $\Delta T_c$ width, at fixed disorder scale, the height, shape and sign of the peak change significantly between configurations (see also the results of configuration B presented before – fig.5), indicating that the effect drastically depends on the local distribution of disorder, making hard to determine the scale of disorder hidden behind the SHE without a prior knowledge of the disorder distribution in the material.

Moreover, the simulation indicates that if $\Delta T_c$ is narrow, as it happens in the $\Delta T_c$=0.1K case, the longitudinal resistivity may even show a double-transition feature, which indicates itself the presence of disorder. In this case, the SHE peak reflects this feature, by showing a double-peak behavior, as it happens for the SHE peaks at 30μm in the configuration A (fig. 6b) and at 70μm in the configuration C (fig. 6d). This does not happen for the broader case $\Delta T_c$=1K, in which neither the longitudinal, neither the transverse signals present a double-transition/double-peak behavior (fig 6e-h).

## DISCUSSION

Following the insights provided by the simulations, a quantitative comparison in terms of absolute values between any experimental results and the simulations would be impractical, but a qualitative comparison is of fundamental importance.

The thin films of Fe(Se,Te) and Nb presented in the previous section are particularly suitable for this kind of comparison, since they present a significant difference in their superconducting transition width values: the critical temperature of Fe(Se,Te) depends on the composition (Se/Te ratio and excess Fe) [39] and, in case of thin films, also on the strain induced by the substrate [40]. These quantities vary on nanometric scale [41]. Therefore, the resistive superconducting transition usually appears smeared. On the other hand, Nb is a pure element superconductor, and usually exhibits a sharp transition. Moreover, the discrepancy between Nb and Fe(Se,Te) has also an intrinsic origin, related to the amplitude of the thermal fluctuations, that can be evaluated by the Ginzburg – Levanyuk (G-L) parameter Gi (eq. 1) [43]. Indeed, Gi ~ $\delta T/T_c$ [40], where $\delta T$ is the range of temperatures where the fluctuations are relevant. For Nb $Gi_{Nb} = 10^{-6} - 10^{-10}$ [42,43], while for Fe(Se,Te) $Gi_{Fe(Se,Te)} \sim 10^{-2} - 10^{-3}$ [43, 44], thus resulting in an intrinsic sharp transition for Nb, and in an intrinsic smeared transition for Fe(Se,Te).

The intrinsic difference of the superconducting transition width between Fe(Se,Te) and Nb enables the correlation with the simulations. In particular, it is reasonable to compare the simulations with $\Delta Tc=1K$ to the experimental data of Fe(Se,Te), and the simulations with $\Delta Tc=0.1K$ with the results of Nb.

In the following we discuss point by point the main results coming by simulation.

**Point 1):** the width of the SHE peak matches the width of the superconducting transition. As already discussed in fig. 2, the width of the SHE peaks of Nb and Fe(Se,Te) reflects the width of the corresponding superconducting transition, which is, for both sample, broader in Hall bar 2 rather than in Hall bar 1. Interestingly, the good match between the experimental measured width of the SHE peaks with the width of the superconducting transition, confirms the outcome of the finite element simulations.

**Point 2):** the amplitude of the SHE peaks (in absolute value) increases when decreasing the transition width $\Delta T_c$.

As discussed before, Fe(Se,Te) and Nb intrinsically present a different width of the superconductive transition, being narrow in Nb and broader in Fe(Se,Te). Finite element simulations revealed that the width of the superconducting transition affects the appearance of the SHE by influencing the amplitude of the peak, which results higher when narrowing $\Delta T_c$. Our experimental data agree with this statement. Indeed, the ratio between the peak amplitude and the corresponding resistivity in Nb and Fe(Se,Te) measurements reveals that the SHE peaks in Nb are one order of magnitude higher than those observed in Fe(SeTe).

**Point 3**): once fixed all the other parameters of the problem, i.e at a fixed disorder configuration, the SHE peak becomes narrower and smaller in magnitude with increasing spatial scale of disorder, as long as the spatial scale of disorder is much smaller than the size of the sample.

Such a situation is way hard to be reproduced experimentally, but it can be mimicked by the application of the magnetic field. Indeed, this is the case of Fe(Se,Te), where the main effect of the magnetic field is the smearing of the superconducting transition, which produces the "homogenization" of the spatial inhomogeneity present in the sample. Indeed, SHE of Fe(Te,Se) in applied magnetic field (Figs. 3d and 3f) exhibits a similar behaviour of SHE varying the disorder scale (Fig.5b).

**Point 4)**: the amplitude, width and sign of the SHE peak depends on the particular random configuration of disorder.

Different types of inhomogeneities lead to various scales of disorder. For example, an accidentally damaged sample may have a large scale of disorder, comparable to the sample's physical dimensions, while compositional and electronic inhomogeneities may occur on much smaller scales. However, the simulations suggest that the SHE depends “randomly” on the local distribution of disorder (fig. 6), making impossible to determine the scale of disorder and its distribution within the sample basing our analysis just on SHE peaks. This characteristic is reproduced experimentally: the peaks drastically change sign and shape in an unpredictable way moving from the two Hall bars of both Fe(Se,Te) and Nb thin films. Also the effect of magnetic field, as observed in Nb thin films (Fig. 3b), can be unpredictable. Indeed, when the main effect of the magnetic field is to shift the transition, as in Nb, when phases with different Hc2 coexist, the evolution of the disorder configuration with temperature may be dramatically different at different magnetic fields. SHE peaks whose magnitude and sign vary with magnetic field in Nb films have been observed also by Sengupta and coworkers [27]. We argue that these signatures of spatial inhomogeneities of electronic and superconducting properties, evidenced by the application of a magnetic field, may originate from superconducting phase inhomogeneity of these samples [45].

**Point 5):** more abrupt spatial variations of Tc are increasingly apparent in Vxx and Vxy with decreasing $\Delta T_c$, possibly even evidencing "double" transitions that are instead averaged out for larger $\Delta T_c$.

The simulations point out that the longitudinal resistivity exhibits a double-transition feature if $\Delta T_c$ is narrow ($\Delta T_c$=0.1K). This happens because, as already pointed out in fig. 4b, narrowing the transition implies a more abrupt change in the resistivity when moving from one region of the sample to another, which implies a higher sensitivity to abruptly changing $T_c$ distribution. On the other hand, when $\Delta T_c$ is larger, it averages the $T_c$ distribution, smearing out any double transition feature in $\rho$xx. It is interesting to note that, independently on the width of $\Delta T_c$, the SHE peak keeps appearing in any case, eventually showing a double-peak feature as pointed out in fig. 6. These observations suggest that the SHE, being a zero-signal, is a very sensitive tool to detect the presence of inhomogeneities, able to recognize it even when the longitudinal resistance in apparently sharp and "blind" to inhomogeneities effects.

The experimental analysis of Nb (sharp transition superconductor) and Fe(Se,Te) (smeared transition superconductor) confirms this statement. Indeed, the Fe(Se,Te) does not show any double-transition feature, neither any double peak in SHE. On the other hand, the superconducting transition across Hall bar 2 of Nb shows a double-transition, which is accurately captured by the SHE, showing a two-peak feature. Moreover, the in-field behaviour of the Nb data revealed that the double-transition feature is suppressed at low field, while the double-peak feature in the corresponding SHE keeps existing up to the highest measured.

**CONCLUSIONS**

In conclusion, our study provides valuable insights into the SHE observed in superconducting materials. We demonstrate that spatial inhomogeneities in the critical temperature play a significant role in the emergence of the SHE. In particular, our findings indicate that the SHE peak is strongly influenced by the width of the superconducting transition, the scale of inhomogeneities in the sample, and the specific random configuration of inhomogeneities within it. By the experimental analysis of the SHE behaviour, together with our finite elements simulations of the phenomena, we establish a correlation between experimental results and simulated scenarios, supporting the hypothesis that spatial inhomogeneity, rather than intrinsic properties of the superconductors, can explain the occurrence of the SHE across different superconducting materials. The superconducting materials under our analysis, Nb and Fe(Se,Te) are particularly suitable for this study, since they belong to very different superconducting families. In particular, Fe(Se,Te) and Nb intrinsically present a different width of the superconductive transition, being narrow in Nb and broader in Fe(Se,Te). Despite the significant discrepancies between Nb and Fe(Se,Te), the SHE analysis on both samples yielded similar results, both of which are consistent with the predictions from the simulations. This allows us to assert that the origin of the SHE can be attributed to the presence of inhomogeneities in the samples, regardless of the type of superconductor being studied.

In addition, our work offers a predictive approach to identify inhomogeneities within superconducting samples. We found that longitudinal resistivity measurements can reveal the presence of disorder by displaying a double transition, but this is true only for superconductors with narrow transition widths, otherwise an "averaging" effect comes out. In contrast, our results demonstrate that the SHE peak consistently appears, regardless of the transition width, and hence on the specific superconductor under study.

This highlights that SHE is an exceptionally sensitive indicator of disorder, offering a robust and straightforward method for detecting inhomogeneities across different superconducting materials, able to recognize it even when the longitudinal resistance in apparently sharp and "blind" to inhomogeneities effects. This is the case of Hall bar 1 of Nb, and of Fe(Se,Te) (see fig. 2). Both the samples could appear very homogeneous looking at the only resistivity curve, but they both show SHE, indicating the presence of disorder.

It is important to point out to that the model we propose, based on inhomogeneities, includes the possibility that these are of electronic nature, for example, related to thermal/magnetic fluctuations. In this sense, our model naturally explains the transversal resistance reported by Sengupta et al. [29]. But, differently from their conclusions that exclude the role of material structure and morphology of the film, claiming an emergent electronic phenomenon, our simulations demonstrate that inhomogeneities of whatever origin, whether structural or electronic, are a certain cause of random current paths, at the basis of SHE.

Finally, we do not exclude that other mechanisms, such as vortex-antivortex motion or time-reversal symmetry breaking, play a role in the raising of a SHE. However, our work points out the necessity of considering case by case the effect of inhomogeneities as an ineluctable background on which other phenomena can overlap. To this scope our analysis suggests some reasonable criteria to disentangle the SHE induced by inhomogeneities, including the correspondence between the width in temperature of the SHE peak and the width of the superconducting transition, even in applied magnetic field.

In conclusion, our results set new constraints on understanding the SHE and highlight the sensitivity of the effect to local variations in superconducting properties, paving the way for further research into the influence of inhomogeneities on superconducting phenomena.

## ACKNOWLEDGMENTS

The authors acknowledge R. Russo for supplying the Nb sample and G. Lamura for fruitful suggestions and discussions. The authors acknowledge PNRR, Missione 4, Componente 2, Avviso 3264/2021 “Innovative Research Infrastructure on applied Superconductivity – IRIS”.

## BIBLIOGRAPHY

[1] Nagaosa, N.; Sinova, J.; Onoda, S.; et al. Anomalous Hall effect. *Rev. Mod. Phys.* **2020**, 82, 1539

[2] Liu, E.; Sun, Y.; Kumar, N.; Muechler, L.; *et al.* Giant anomalous Hall effect in a ferromagnetic kagome-lattice semimetal. *Nature Phys* **2018**, 14, 1125–1131

[3] Nagaosa, N.; Anomalous Hall Effect – A New Perspective. *J. Phys. Soc. Jpn*. **2006**, 75, 042001

[4] Smejkal, L.; MacDonald, A. H.; Sinova, J.; et al. Anomalous Hall antiferromagnets. *Nature Review Materials* **2022**, 7, 482-496

[5] Nakatsuji, S.; Kiyohara, N.; Higo, T. Large anomalous Hall effect in a non-collinear antiferromagnet at room temperature. *Nature* **2015**, 527, 212–215

[6] Ikhlas, M.; Tomita, T.; Koretsune, T.; et al. Large anomalous Nernst effect at room temperature in a chiral antiferromagnet. *Nature Phys* **2017**, 13, 1085–1090

[7] Sakai, A.; Mizuta, Y. P; Nugroho, A. A.; et al. Giant anomalous Nernst effect and quantum-critical scaling in a ferromagnetic semimetal. *Nat. Phys.* **2018,** 14, 1119–1124

[8] Caglieris, F.; Wuttke, C.; Sykora, S.; et al. Anomalous Nernst effect and field-induced Lifshitz transition in the Weyl semimetals TaP and TaAs. *Phys. Rev. B* **2018**, 8, 201107

[9] Wuttke, C.; Caglieris, F.; Sykora, S.; et al. Berry curvature unravelled by the anomalous Nernst effect in Mn3Ge. *Physical Review B* **2019**, 100, 085111

[10] Ceccardi, M.; Zeugner, A.; Folkers, L.C.; et al. Anomalous Nernst effect in the topological and magnetic material $MnBi_4Te_7$. *npj Quantum Mater.* **2023**, 8, 76

[11] Wang, X.; Vanderbilt, D.; Yates, J. R.; et al. Fermi-surface calculation of the anomalous Hall conductivity. *Physical Review B* **2007**, 76, 195109

[12] Haldane, F. D. M. Berry Curvature on the Fermi Surface: Anomalous Hall Effect as a Topological Fermi-Liquid Property. *Phys. Rev. Lett.* **2004**, 93, 206602

[13] Smit, J. The spontaneous hall effect in ferromagnets I; Physica, Volume 21, Isuues 6-10, **1955**; pp 877–887.

[14] Smit, J. The spontaneous hall effect in ferromagnets II; Physica, Volume 24, Isuues 1-5, **1958**; pp 39-51.

[15] Berger, L. Side-Jump Mechanism for the Hall Effect of Ferromagnets. *Phys. Rev. B* **1970**, 2, 4559

[16] da Luz, M. S.; de Carvalho Jr., F. J. H.; dos Santos, C. A. M.; et al. Observation of asymmetric transverse voltage in granular high-$T_c$ superconductors. *Physica C* **2005**, 419, 71-78

[17] da Luz, M. S.; dos Santos, C. A. M.; Shigue, C. Y.; et al. *Physica C* **2009**, 469, 60-63

[18] Vasek, P.; Shimakage, H.; Wang, Z. Transverse voltage in zero external magnetic

fields, its scaling and violation of the time-reversal symmetry in $MgB_2$. *Physica C* **2004**, 411, 164-169

[19] Sengupta, S.; Monteverde, M.; Loucif, S.; et al. Spontaneous voltage peaks in superconducting Nb channels without engineered asymmetry. *Phys. Rev. B* **2024**, 109, L060503

[20] Xu, X.; Li, Y. & Chien, C.L. Anomalous transverse resistance in the topological superconductor $\beta$-$Bi_2Pd$. *Nat Commun* **2022**, 13, 5321

[21] Shipulin, I.; Stegani, N.; Maccari, I.; et al. Calorimetric evidence for two phase transitions in $Ba_{1-x}K_xFe_2As_2$ with fermion pairing and quadrupling states. *Nat Commun* **2023**, 14, 6734

[22] Chen, L.; Xiang, Z.; Tinsman, C.; et al. Spontaneous Nernst effect in the iron-based superconductor Fe1+$y$Te1$-x$Se$x$. *Phys. Rev. B* **2020**, 102, 054503

[23] Li, L.; Alidoust, N.; Tranquada, J. M.; et al. Unusual Nernst Effect Suggesting Time-Reversal Violation in the Striped Cuprate Superconductor La2$-x$Ba$x$CuO4. *Physical Review Letters* **2011**, 107, 277001

[24] Vaglio, R.; Attanasio, C.; Maritato, L.; et al. Explanation of the resistance-peak anomaly in nonhomogeneous superconductors. *Phys. Rev. B* **1993**, 47, 15302

[25] Lv, Y.; Dong, Y.; Lu, D.; et al. Anomalous transverse resistance in 122-type iron-based superconductors. *Sci Rep* **2019**, 9, 664

[26] Segal, A.; Karpovski, M.; Gerber, A. Inhomogeneity and transverse voltage in superconductors. *Phys. Rev. B* **2011**, 83, 094531

[27] Siebold, T.; Carballeira, C.; Mosqueira, J.; et al. Current redistributions in superconductors with non-uniformly distributed Tc-inhomogeneities. *Physica C* **1997**, 282-287, 1181-1182

[28] Sengupta, S.; Farhadizadeh, A.; Youssef, J.; et al. Transverse resistance due to

electronic inhomogeneities in superconductors. arXiv:2407.16662 [cond-mat.supr-con]

[29] Zhao, D.; Cui, W.; Gong, G.; et al. Electronic inhomogeneity and phase fluctuation in one-unit-cell FeSe films. *Nat. Commun.* **2024**, 15, 3369

[30] Miyai, Y.; Ishida, S.; Ozawa, K.; Yoshida, Y.; Eisaki, H.; Shimada, K.; Iwasawa, H. Visualization of spatial inhomogeneity in the superconducting gap using micro-ARPES. *Science and Technology of Advanced Materials* **2024**, 2379238

[31] Iebole, M.; Braccini, V.; Bernini, C.; et al. Fe(Se,Te) Thin Films Deposited through Pulsed Laser Ablation from Spark Plasma Sintered Targets. *Materials* **2024**, *17*(11), 2594

[32] Piperno, L.; Vannozzi, A.; Augieri, A.; et al. High-performance Fe(Se,Te) films on chemical $CeO_2$-based buffer layers. *Sci Rep.* **2023**, 13, 569

[33] Palenzona, A.; Sala, A.; Bernini, C.; et al. A new approach for improving global critical current density in $Fe(Se_{0.5}Te_{0.5})$ polycrystalline materials. *Supercond. Sci. Technol.* **2012**, 25, 115018

[34] Piperno, L.; Vannozzi, A.; Pinto, V.; et al. Chemical CeO2-Based Buffer Layers for Fe(Se,Te) Films. *IEEE TRANSACTIONS ON APPLIED SUPERCONDUCTIVITY* **2022**, 32, 4, 7300205

[35] Russo, R.; Cianchi, A.; Akhmadeev, Y. H.; et al. UHV arc for high quality film deposition. *Surface & Coatings Technology* **2006**, 201, 3987-3992

[36] Russo, R.; Catani, L.; Cianchi, A.; et al. High quality superconducting niobium films produced by an ultra-high vacuum cathodic arc. *Supercond. Sci. Technol.* **2005** 18, L41-L44

[37] **Supplemental Materials**

[38] A.Maljuk, S.Watauchi, I.Tanaka and H.Kojima. The effect of $B_2O_3$ addition on $La_{2-x}Sr_xCuO_4$ single-crystal growth *Journal of Crystal Growth*, volume 212, Issues 1–2,

**2000**, Pages 138-141.

[39] Zhuang, J.; Yeoh, W.; Cui, X.; et al. Unabridged phase diagram for single-phased $FeSe_xTe_{1-x}$ thin films. *Sci Rep* **2014**, 4, 7273

[40] Bellingeri, E.; Kawale, S. ; Pallecchi, I. ; et al. Strong vortex pinning in FeSe0.5Te0.5 epitaxial thin film. *Appl. Phys. Lett.* 100, 082601 **(2012)**

[41] Scuderi, M.; Pallecchi, I.; Leo, A.; et al. Nanoscale analysis of superconducting Fe(Se,Te) epitaxial thin films and relationship with pinning properties. *Sci Rep* **2021**, 11, 20100

[42] Hunte, F.; Jaroszynski, J.; Gurevich, A.; et al. Two-band superconductivity in LaFeAsO0.89F0.11 at very high magnetic fields. *Nature* **2008**, 453, 903–905

[43] Putti, M.; Pallecchi, I.; Bellingeri, E.; et al. New Fe-based superconductors: properties relevant for applications. *Supercond. Sci. Technol.* **2010**, 23**,** 034003

[44] Kazumasa, I.; Hanisch, J.; Tarantini, C. Fe-based superconducting thin films on metallic substrates: Growth, characteristics, and relevant properties. *Appl. Phys. Rev.* **2018**, 5, 031304

[45] Durkin, M.; Garrido-Menacho, R.; Gopalakrishnan, S.; et al. Rare-region onset of superconductivity in niobium nanoislands. *Phys. Rev. B* **2020**, 101, 035409

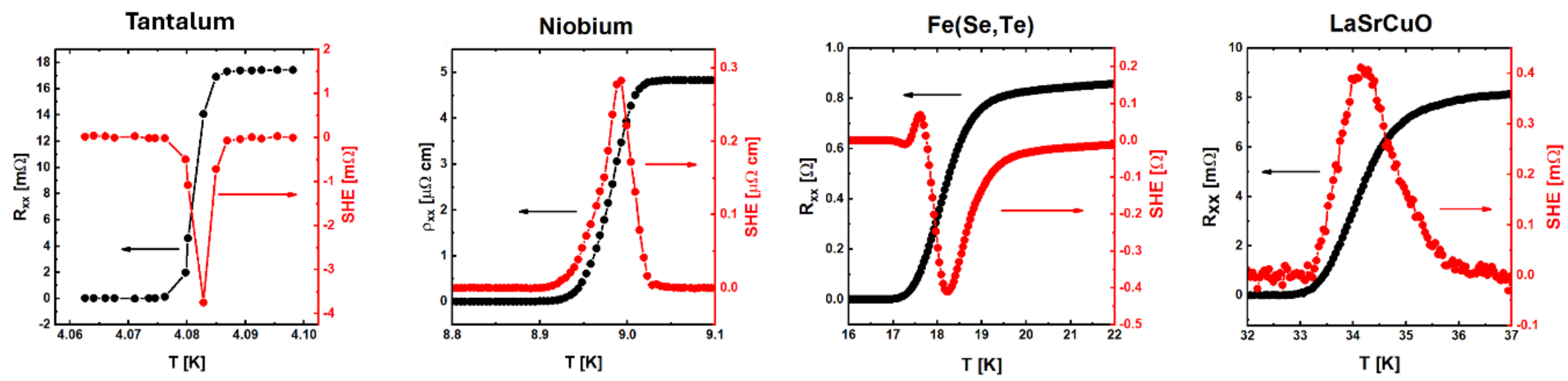

**Figure 1**: SHE appearing in: (a) 80 nm thick Ta film, (b) 80 nm thick Nb film, (c) 100 nm thick Fe(Se,Te) film, and (d) bulk LSCO crystal.

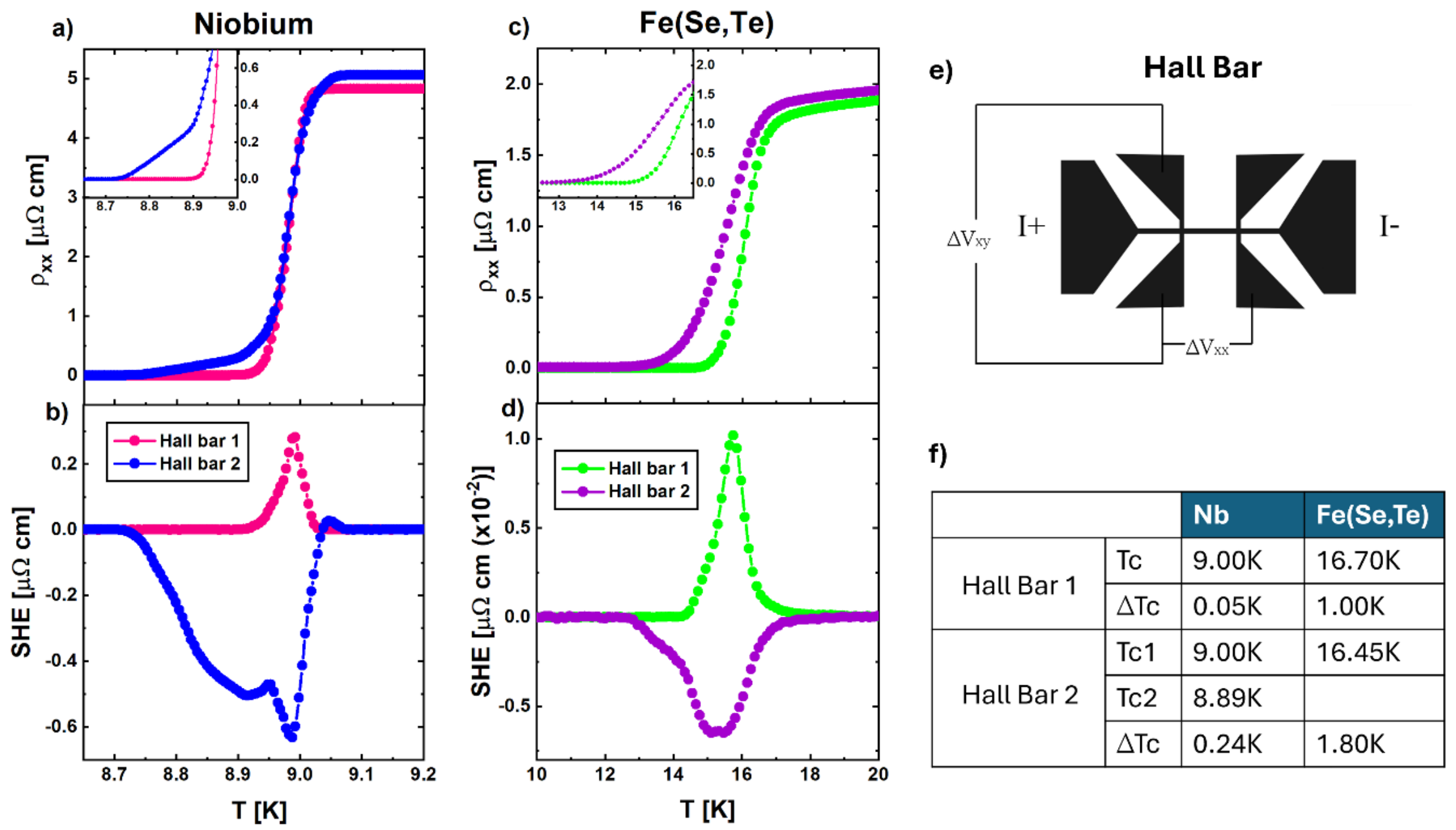

| | | Nb | Fe(Se,Te) |
|---|---|---|---|
| Hall Bar 1 | Tc | 9.00K | 16.70K |
| | ΔTc | 0.05K | 1.00K |
| Hall Bar 2 | Tc1 | 9.00K | 16.45K |
| | Tc2 | 8.89K | |
| | ΔTc | 0.24K | 1.80K |

**Figure 2**: Longitudinal resistivity (a-c) and SHE (b-d) of the Nb thin film and Fe(Se,Te) thin film respectively, belonging to Hall bar 1 and Hall bar 2. Inset: enlarged view of the feet of the longitudinal transitions ρxx. Panel e: scheme of the Hall bar. Panel f: table of Tc and ΔTc (transition width) for both Hall bars of Nb and Fe(Se,Te).

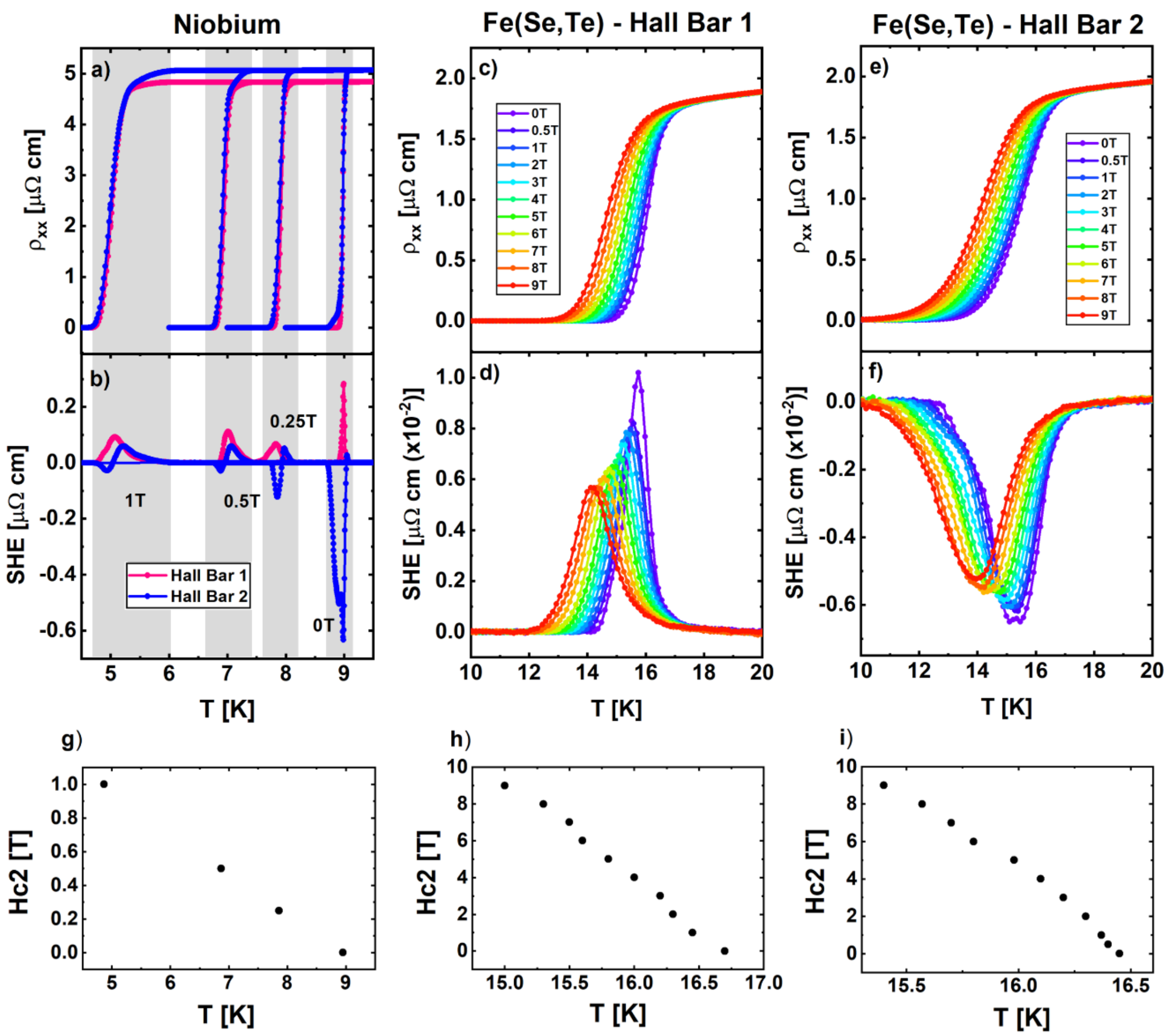

**Figure 3**: Longitudinal resistivity and SHE of: two Hall bars of Nb thin film measured at 1T, 0,5T, 0,25T, 0T (panels a,b); Hall Bar 1 of Fe(Se,Te) measured up to 9T (panels c,d); Hall Bar 2 of Fe(Se,Te) measured up to 9T (panels e,f). Panels g-f: Upper critical field Hc2 of Niobium, Fe(Se,Te) – Hall Bar 1 and Fe(Se,Te) – Hall Bar 2, respectively.

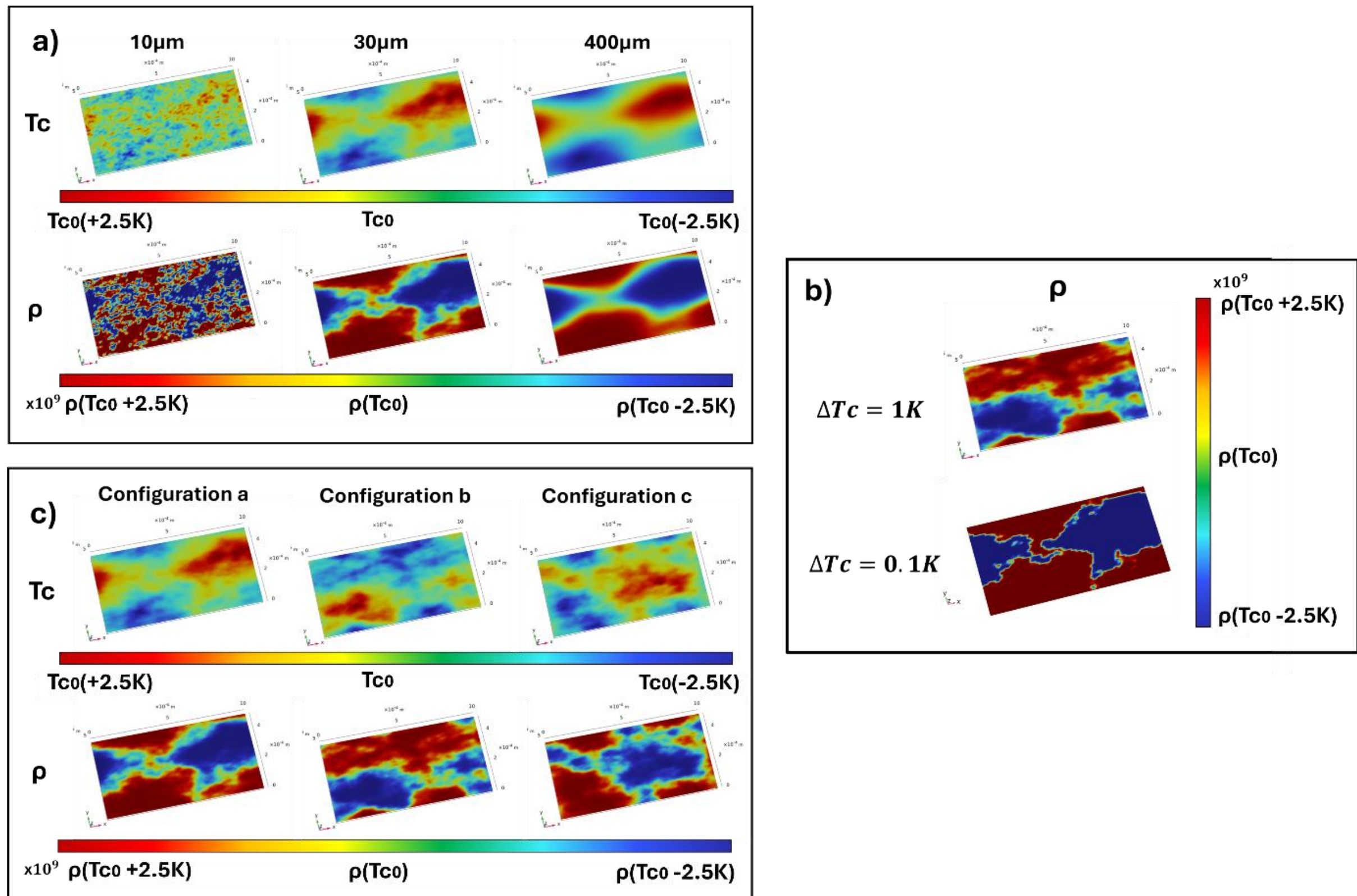

**Figure 4**: (a): 2D maps of $T_c$ and resistivity for $\Delta T_c$=1 K, for 10, 30 and 400 micrometers simulated values of the spatial scale of disorder, for the particular disorder configurations a, (b) 2D maps of resistivity for $\Delta T_c$=1 K and $\Delta T_c$=0.1 K, for disorder scale 30 micrometers, in the disorder configuration a. (c) 2D maps of $T_c$ and resistivity for $\Delta T_c$=1 K and disorder scale 30 micrometers, for the 3 disorder configurations a, b and c.

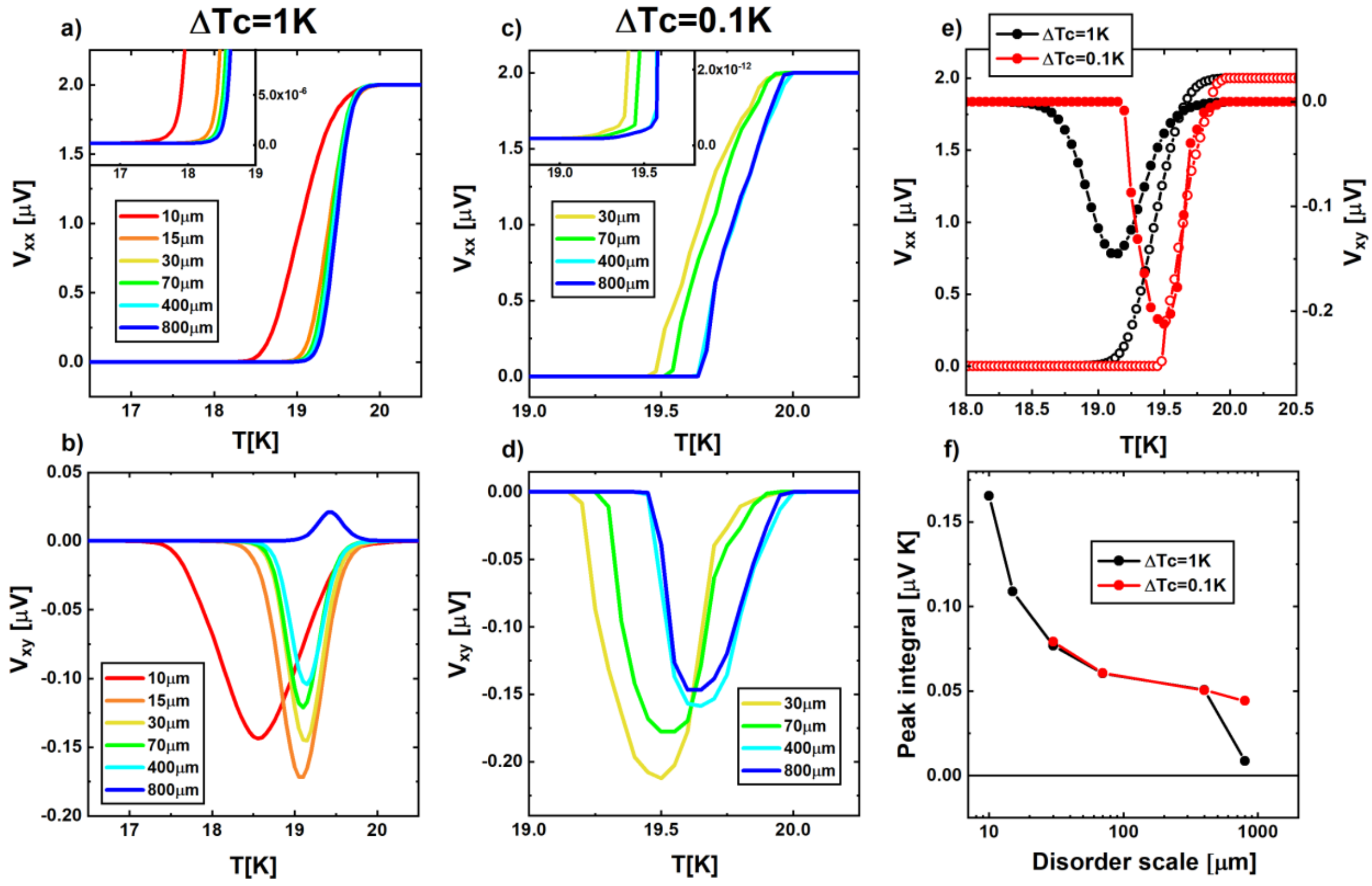

**Figure 5**: (a): Longitudinal voltage drop Vxx as a function of temperature for $\Delta T_c$=1 K for all the 6 simulated scales of disorder. (b): Transverse voltage drop Vxy as a function of temperature for $\Delta T_c$=1 K for all the 6 simulated scales of disorder. (c): Longitudinal voltage drop Vxx as a function of temperature for $\Delta T_c$=0.1 K for the 4 simulated scales of disorder. (d): Transverse voltage drop Vxy as a function of temperature for $\Delta T_c$=0.1 K for the 4 simulated scales of disorder. (e): Comparison of the SHE peaks (right-y axes) and the longitudinal voltage drop Vxx (left-y axes) between the simulation with $\Delta T_c$=1 K (black empty and full dots) and the simulation with $\Delta T_c$=0.1 K (red empty and full dots), for a disorder scale of 30μm. (f): Plot of the absolute value of the peak integral as a function of the disorder scale, both for the $\Delta T_c$=1 K and $\Delta T_c$=0.1 K cases.

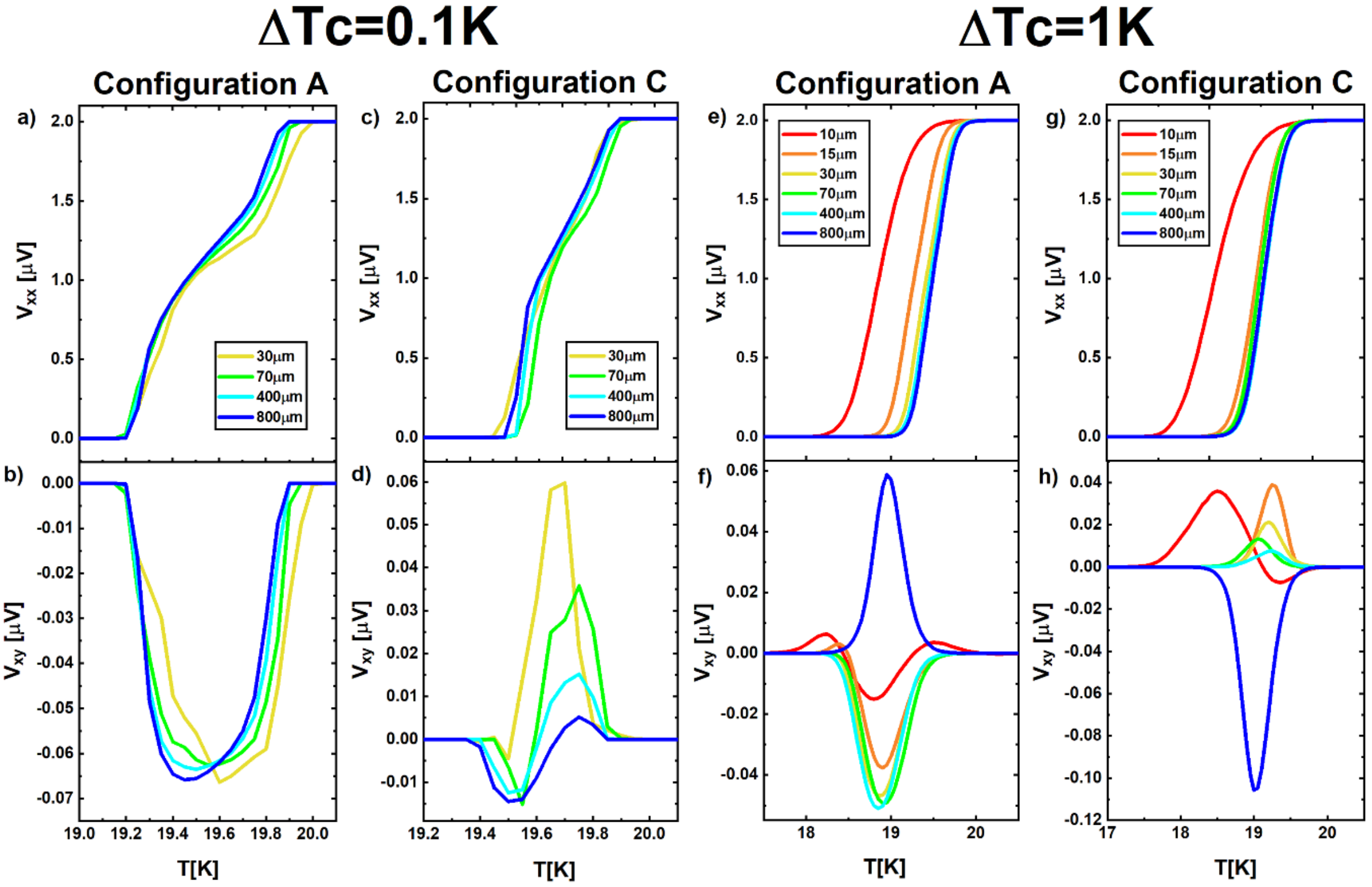

**Figure 6**: Longitudinal Vxx and transverse Vxy voltage drops as a function of temperature for $\Delta T_c$=0.1 K and $\Delta T_c$=1 K for all the 6 simulated scales of disorder, referring to the configuration of disorder A and configuration of disorder C. In particular: Configuration A - $\Delta T_c$=0.1 K (panels a, b); Configuration C - $\Delta T_c$=0.1 K (panels c, d); Configuration A - $\Delta T_c$=1 K (panels e, f); Configuration C - $\Delta T_c$=1 K (panels g, h).